Title

**The spatial distribution of impact craters on Ryugu**

Authors

**Naoyuki Hirata [a, *], Tomokatsu Morota [b], Yuichiro Cho [b], Masanori Kanamaru [c], Sei-ichiro Watanabe [d], Seiji Sugita [b], Naru Hirata [e], Yukio Yamamoto [f], Rina Noguchi [f], Yuri Shimaki [f], Eri Tatsumi [b], Kazuo Yoshioka [g], Hirotaka Sawada [f], Yasuhiro Yokota [f], Naoya Sakatani [f], Masahiko Hayakawa [f], Moe Matsuoka [f], Rie Honda [h], Shingo Kameda [i], Mamabu Yamada [j], Toru Kouyama [k], Hidehiko Suzuki [l], Chikatoshi Honda [e], Kazunori Ogawa [a], Yuichi Tsuda [f], Makoto Yoshikawa [f], Takanao Saiki [f], Satoshi Tanaka [f], Fuyuto Terui [f], Satoru Nakazawa [f], Shota Kikuchi [f], Tomohiro Yamaguchi [f], Naoko Ogawa [f], Go Ono [m], Yuya Mimasu [f], Kent Yoshikawa [m], Tadateru Takahashi [f], Yuto Takei [f], Atsushi Fujii [f], Hiroshi Takeuchi [f], Tatsuaki Okada [f], Kei Shirai [f], Yu-ichi Iijima [n].**

* Corresponding Author E-mail address: hirata@tiger.kobe-u.ac.jp

**Authors' affiliation**

[a] Graduate School of Science, Kobe University, Rokkodai, Kobe 657-8501, Japan.

[b] Graduate School of Science, University of Tokyo, Tokyo 113-0033, Japan.

[c] Department of Earth and Space Science, Osaka University, Osaka, Japan.

[d] Graduate School of Environmental Studies, Nagoya University, Nagoya 464–8602, Japan.

[e] University of Aizu, Aizu-Wakamatsu, Japan.

[f] Institute of Space and Astronautical Science, JAXA, Sagamihara, Japan.

[g] Graduate School of Frontier Sciences, University of Tokyo, Kashiwa, Japan.

[h] Kochi University, Kochi, Japan.

[i] College of Science, Rikkyo University, Tokyo, Japan.

[j] Planetary Exploration Res. Center, Chiba Inst. Tech., Chiba, Japan.

[k] National Institute of Advanced Industrial Science and Technology, Tokyo, Japan.

[l] Meiji University, Tokyo, Japan.

[m] Research and Development Directorate, JAXA, Sagamihara, Japan.

[n] Deceased

**Proposed Running Head:** Craters on Ryugu

Editorial Correspondence to:
Dr. Naoyuki Hirata
Kobe University, Rokkodai 1-1 657-0013
Tel/Fax +81-7-8803-6566

**Highlights**

- We examined the spatial distribution of impact craters on Ryugu
- We completed a global impact crater catalogue of Ryugu (D>20 m)
- Crater density variations cannot be explained by the randomness of cratering
- More craters are seen at lower latitudes and less at higher latitudes
- There are fewer craters in the western bulge and more around the meridian

**Abstract**

Asteroid 162173 Ryugu has numerous craters. The initial measurement of impact craters on Ryugu, by Sugita et al. (2019), is based on Hayabusa2 ONC images obtained during the first month after the arrival of Hayabusa2 in June 2018. Utilizing new images taken until February 2019, we constructed a global impact crater catalogue of Ryugu, which includes all craters larger than 20 m in diameter on the surface of Ryugu. As a result, we identified 77 craters on the surface of Ryugu. Ryugu shows variation in crater density which cannot be explained by the randomness of cratering; there are more craters at lower latitudes and fewer at higher latitudes, and fewer craters in the western bulge (160°E – 290°E) than in the region around the meridian (300°E – 30°E). This variation implies a complicated geologic history for Ryugu. It seems that the longitudinal variation in crater density simply indicates variation in the crater ages; the cratered terrain around the meridian seems to be geologically old while the western bulge is relatively young. The latitudinal variation in crater density suggests that the equatorial ridge of Ryugu is a geologically old structure; however, this could be alternatively explained by a collision with many fission fragments during a short rotational period of Ryugu in the past.

## 1. Introduction

Ryugu is a top-shaped asteroid with a mean radius of 448 m and a rotational obliquity of 8 degrees (Watanabe et al., 2019). Since JAXA's Hayabusa2 spacecraft arrived at asteroid 162173, Ryugu, on June 27, 2018, the onboard optical navigation cameras (ONC) have obtained numerous images of Ryugu and revealed many surface features, including abundant impact craters (Sugita et al., 2019). This cratering suggests that a major process in the formation of the surface of Ryugu has been via impacts from other bodies. The low bulk density of Ryugu implies that the asteroid is composed of re-accumulated unconsolidated fragments that probably resulted from impact during a process such as catastrophic disruption (Watanabe et al., 2019). The initial report of impact craters on Ryugu (Sugita et al. 2019) was produced via the utilization of ONC images from July to August 1, 2018, which identified approximately 30 impact craters from the limited coverage of Ryugu. The surface crater retention age was estimated to be in the order of $10^7$ or $10^8$ years, based on the number density of craters of 100 – 200 m in diameter (D). In addition, Sugita et al. (2019) reported that the crater size-frequency distribution (CSFD) of Ryugu shows a lack of small craters (D < 100 m), which indicates that the average resurfacing of the top ~1-meter layer on Ryugu takes less than $10^6$ years. Ryugu has a west/east dichotomy (Sugita et al. 2019); its western side, the so-called western bulge, (160°E – 290°E), has a high albedo, a low number density of large boulders, is topographically high, and has a bluish color as compared to the eastern side. The equatorial ridge located in the eastern hemisphere is slightly offset towards the south.

The main purpose of this paper is (i) to present the basic information (the location and size of each craters), which is not listed in the initial report by Sugita et al. (2019), (ii) to accomplish a global catalogue of all the craters ($D \geq 10$ - 20 m), utilizing additional images acquired after the publication by Sugita et al. (2019) which have allowed the investigation of craters over the entire surface of Ryugu, (iii) to investigate the statistical significance of the spatial distribution of the impact craters. Note that we do not discuss the crater size-frequency distribution or the depletion and retention time of small craters, as these topics will be included in Morota et al. (in preparation).

## 2. DATA and Method

### 2.1. Crater counting

This study utilized 340 ONC images for crater counting (Table 1): (i) 96 images obtained in July 20 2018 with a ground resolution of 0.72 m/pixel providing global coverage excluding the polar regions > 70 degrees, (ii) 85 images obtained on August 1 2018 with a resolution of 0.69-0.55 m/pixel, providing regional coverage of low latitudes < 40 degrees, (iii) 11 images obtained on October 4 2018 with a resolution of 0.32 m/pixel, providing coverage of the north pole, (iv) 18 images on October 30 2018 with a resolution of 0.62 m/pixel providing coverage of the north and south poles (v) 26 images obtained on February 28 2019 with a resolution of 0.67 m/pixel providing coverage of the north and south poles, (vi) 52 images obtained on August 23 2018 with a resolution of 2.6 m/pixel providing coverage of the south pole, (vii) 52 images obtained on January 24 2019 with a resolution of 2.1 m/pixel providing coverage of the north pole. Because the Hayabusa 2 spacecraft is generally remaining in the same position above the sub-Earth point of the asteroid's surface and the rotational obliquity of Ryugu is small, the emission angles of the images (i)-(v) are large for polar regions. The low-emission images of Ryugu's polar regions, (vi) and (vii), were obtained at a position distant from the equatorial plane. All of the impact craters were identified from these ONC images. These ONC images will be freely available at the end of 2019. Craters with diameters larger than 10-20 m could be clearly identified as these images had a resolution of at least 10 pixels. Also, the images enabled us to identify all impact craters exceeding 10 – 20 meters diameter over the entire surface of Ryugu. Therefore, to ensure that crater counting was not affected by image resolution, D = 20 m was set as the minimum diameter threshold.

We identified all circular or quasi-circular depressions as craters for this study. This included cases where the circular depression lacked a raised rim. However, we did not regard circular features without topographic depressions as craters. Based on these criteria, we classified all the candidate craters into types: I-IV (as summarized in Table 2). We judged those classified as types I-III to be distinct crater candidates, and classification IV to be less-distinct crater candidates. For this study, we assumed that type IV phenomena are not craters, and therefore these were not included in crater density and statistical analysis. Features such as topographic depression were judged from shading, stereo pairs, and the shape model. Nonetheless, many craters were more or less degraded and infilled with regolith or boulders and often lacked distinct shape, and this interpretation was therefore often ambiguous. All candidate craters are revealed in Figure 1. Nos. 1- 77 in Figure 1 have been interpreted as distinct craters, and Nos. 78-86 are less-distinct craters. The interpretation of these images varied among researchers.

To measure the size, latitude, and longitude of each crater we utilized the Small Body Mapping Tool (Kahn et al., 2011), a global image mosaic map of Ryugu, and a shape model of Ryugu. The global image mosaic map was built from the ONC images, and the shape model was derived from Watanabe et al. (2019). The Small Body Mapping Tool was used to measure the diameters and locations of the craters for the study. This software enabled measurement of the centers and diameters of craters based on three points selected along the crater rim from a global mosaic map rendered onto the shape model.

A fairly accurate surface area is required to obtain the number density of the craters. In the case of an irregular-shaped asteroid, the definition of the surface area is complicated. When defining it from a shape model, the surface area increases infinitely as the resolution of the shape model increases. To simplify, the surface area was not determined as the total area of the surface polygons composing the shape model, but was instead calculated as the surface of a sphere with a radius of 448 m. The total surface area of Ryugu was thus calculated to be 2.5 $km^2$.

### 2.2. The statistical analysis of the spatial distribution

A statistical test was then performed to evaluate the significance of variations in the crater density. The nearest-neighbor analysis was thus used to discover whether the variation can be explained merely by randomness, following the methodology of Squyres et al. (1997). The distance between each point (i.e. the center of a crater) and its nearest neighboring point (i.e. the center of the nearest neighbor crater) was determined and averaged, and the mean distance observed was compared with the mean distance expected under random distribution. If the observed value was significantly smaller (greater) than the expected value, the distribution was considered to be significantly clustered (ordered) and not random. This method has been applied to evaluate the distribution of craters on various solar system objects.

Phillips et al. (1992) found that, although Venus has a variation in crater density, the variation cannot be distinguished from a completely random distribution; therefore, the variation can be explained by the randomness of the crater production on that planet. Squyres et al. (1997) found that the crater distributions on Callisto and Rhea are not random but are significantly ordered. The spatial distribution of observed craters in heavily cratered terrain transition from random to ordered because craters that form in sparse areas obliterate the relatively few existing craters, filling in the spaces, whereas craters that form in areas of existing crater clustering obliterate the existing craters and reduce the clustering (Lissauer et al. 1988; Squyres et al. 1997). Moreover, partial resurfacing such as lava flow or the mass-movement of rock tends to create clustered crater distribution.

Following Squyres et al. (1997), the $Z$ value was determined to assess the degree by which the value of the observed mean distance ($d_{obs}$) deviates from randomness:

$$Z \equiv \frac{d_{obs}-d_{exp}}{\sigma}, \quad (1)$$

where $d_{exp}$ is the expected mean distance when the distribution is completely random and $\sigma$ is the standard deviation of $d_{exp}$. A positive $Z$ value indicates that the distribution of points is ordered, a negative $Z$ value indicates that it is more clustered, and a value of $Z$ close to 0 indicates that the distribution cannot be distinguished from random. A value for $Z$ was obtained for the distribution of all craters that exceeded a given diameter. The $Z$-statistic was not calculated in cases where the number of craters was lower than $n$ = 2. The observed distance between two points was determined on the basis of the great-circle distance between the two points on the unit sphere, although the shape of Ryugu is not spherical. Values for $d_{exp}$ and $\sigma$ were calculated numerically using the Monte Carlo simulation, following Hirata (2017). This study assumed that (1) $n$ points were produced on the unit sphere, (2) these points were randomly generated because the impactors were assumed to have come from all directions, (3) the mean distance to the nearest neighbor ($d_i$ in $i^{th}$ trial) was measured using the average of the lengths defined by the great-circle distance between two points, (4) based on these three assumptions, 10000 trials were performed, and (5) the average and deviation of $d_i$ ($i$ = 1,2,3, …10000) were obtained as $d_{exp}$ and $\sigma$, respectively.

## 3. Results

On the surface of Ryugu, 77 craters were identified (Fig. 1); the diameters and locations of these craters are listed in Table 3. Figure 2 shows the global distribution of craters that are bordered with rims as lines on the global mosaic map projected in a simple cylindrical projection. We again confirmed a lack of small craters (20 m < D < 100 m) relative to the number of large craters, as has also been found on other small asteroids such as Itokawa, Eros, and Bennu (Thomas and Robinson, 2005; Michel et al. 2009; Walsh et al. 2019). The largest crater on Ryugu is Urashima, at 290 m in diameter, which is equivalent to 32% of the diameter of Ryugu. The second and third largest craters are Cendrillon (224 m) and Kolobok (221 m), respectively. Only these three craters exceed 200 meters in diameter. The average crater density over Ryugu is 4.4 craters/km$^2$ for D ≥ 100 m and 25.0 craters/km$^2$ for D ≥ 20 m.

The distribution of craters on Ryugu is variable, with differences in crater density depending on longitude and latitude (Fig 3a, b). The $Z$ value (Fig. 3c) suggests that the craters on Ryugu are spatially distributed in clusters, in a statistically significant manner. Therefore, the spatial distribution of craters on Ryugu cannot be explained by the randomness of cratering itself. There are more craters at lower latitudes and fewer craters at higher latitudes (Fig. 3b). Polar regions of greater than 40 degrees latitude have only a few craters. There are more craters around the meridian (300°E-30°E), and fewer at the western bulge (160°E – 290°E). Because of this we termed the region between 300°E-30°E and 40°S-40°N the cratered terrain. The crater density in this region is 57.3 craters/km$^2$ (D ≥ 20 m), twice the global average. However, as shown in Fig 3a, the crater density of the western bulge is roughly half the global average. Although less-cratered terrain is often found both surrounding and on the inside of large craters such as Shoemaker crater on Eros (Robinson et al. 2002), we could not identify any landform that could be associated with a putative large crater at either the polar regions or the western bulge of Ryugu.

Ryugu has 11 craters (D ≥ 100 m), of which 5 cut the crest of the equatorial ridge, Ryujin Dorsum. Note that we concluded that it is unlikely that some of the impact craters found on this ridge did not originate from impacts (i.e., the ridge is a landform made by another mechanism such as mass-movement). This is because the large craters cutting across the ridge (No. 1, 3, 4, 6, 8 in Fig. 1) are all clear and circular bowl-shaped depressions and appear to be distinct impact craters, as discussed by Sugita et al. (2019). We therefore performed a statistical test to evaluate the statistical significance of the concentration of craters on the equatorial ridge, Ryujin Dorsum. Using the Monte Carlo method, we estimated the number of craters expected across the equator under random distribution. In detail: (1) the 11 craters exceeding diameters of 100 m, are placed at random points on a sphere with a radius of 448 m using spherical uniform random numbers, (2) the number of craters across the equator is counted, and (3) the average value of the number and its dispersion are obtained from 1000 trials. Results indicate that the expected number of large craters across the equator under random distribution is 2.1 with a dispersion of 1.3. We therefore concluded that the number of observed craters found on the equator (5 craters) was greater than that expected randomly.

## 4. Discussion

### 4.1. Longitude variation

Possible explanations for the longitudinal variation in crater density are: (i) resurfacing from processes such as seismic shaking is inert in the cratered terrain and is more active in the western bulge, (ii) differences in the physical terrain inhibit the formation of an impact crater (e.g., Güttler et al. (2012) proposed an armoring effect owing to the presence of boulders), and (iii) the cratered terrain around the meridian (300°E -30°E) is geologically old, while the formation of the western bulge was relatively recent. Seismic shaking or armoring effects, as in (i) and (ii), should theoretically be more effective for smaller craters; however, even large craters on Ryugu show longitudinal variation which indicates that (i) and (ii) are unlikely. We propose that (iii) is the most likely scenario. This is because the western bulge is a geologically distinct terrain, as mentioned in Section 1. The western bulge is proposed to have been responsible for the deformation of Ryugu, via a process that occurred on only the western side of the asteroid during a short rotational period of Ryugu in the past, while the eastern hemisphere was left structurally intact (Hirabayashi et al. 2019). If so, the crater density of the western bulge could simply be explained by the timing of the short rotational period of Ryugu. Scheeres (2015) proposed that regolith landslides occur on the surface of rapidly spinning asteroids, so the occurrence of such a landslide from the polar regions toward the western hemisphere could be taken into consideration as an alternative explanation for the formation of the western bulge (even though there is no strong evidence that this occurred). Based on the above scenario, the crater density of the western bulge represents the timing of the short rotational period of Ryugu, while the cratered terrain around the meridian was already an old surface when the western bulge formed.

### 4.2. Latitude variation

The common assumption (a cratered terrain is geologically old while a less-cratered terrain is young) leads to the suggestion that the cratered ridge must simply be a geologically old structure. However, we consider that the formation of the equatorial ridge should have caused a resurfacing of the asteroid, thus eliminating the older craters at low latitudes. An interesting cratered equatorial ridge has also been identified on Bennu, (Walsh et al. 2019). Alternatively, the formation of the equatorial ridge of a top-shaped asteroid may coincide with crater formation on the equator due to the occasional collision with rotationally fissioned small satellites. Such collisions are predicted to occur due to the BYORP effect or chaotically dynamic processes (e.g. Jacobson and Scheeres, 2011). If this is the case, the cratered equatorial ridge would not be considered old.

## 5. Conclusion

We have identified 77 craters on the surface of Ryugu. The spatial distribution of the craters on Ryugu is not random, with variations in crater density that can be linked to latitude and longitude; more craters are seen at lower latitudes than at higher latitudes, and there are more craters in the region around the meridian than in the western bulge. The longitudinal variation in crater density could be a possible result of the formation of the western bulge, which is thought to have formed later than the rest of the asteroid due to its lower proportion of craters. The latitude variation in crater density indicates that the equatorial ridge is a fossil structure formed during the short rotational period in the distant past. Other possible explanations for the latitudinal variation are viable, including the notion that it was formed due to frequent collisions with fissioned fragments that fell onto the equatorial ridge during a short rotational period of Ryugu.

**Acknowledgements**

This study was supported by the JSPS International Planetary Network. Images obtained by Hayabusa2 will be freely available via Data ARchives and Transmission System (DARTS) at ISAS/JAXA at the end of 2019 (https://www.darts.isas.jaxa.jp/index.html.en). The shape model of Ryugu has already been released in DARTS. The Small Bodies Mapping Tool has been released at http://sbmt.jhuapl.edu/index.html.

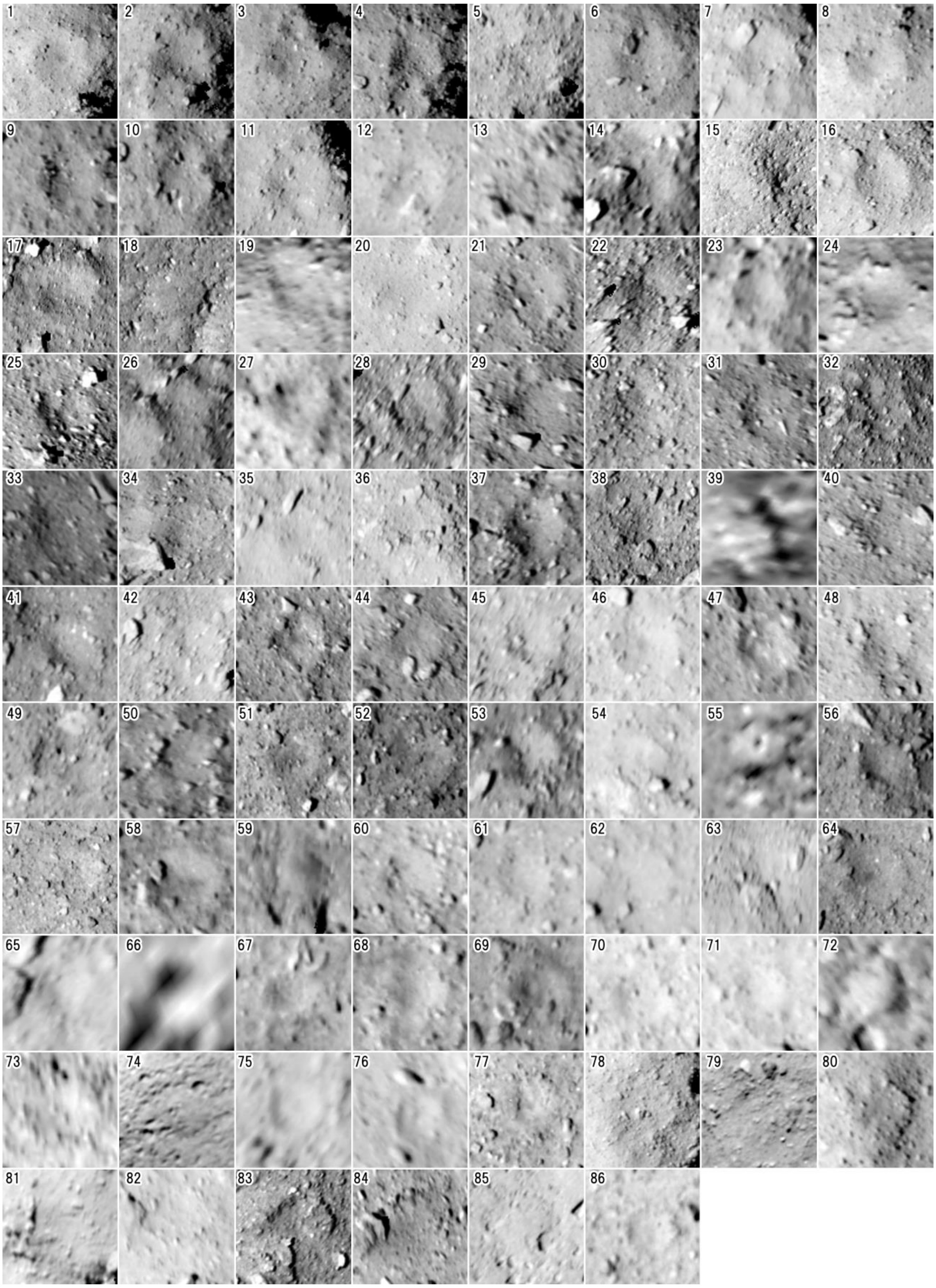

**Figure 1.** Candidate craters that were identified. Numbers correspond to those in Table 3.

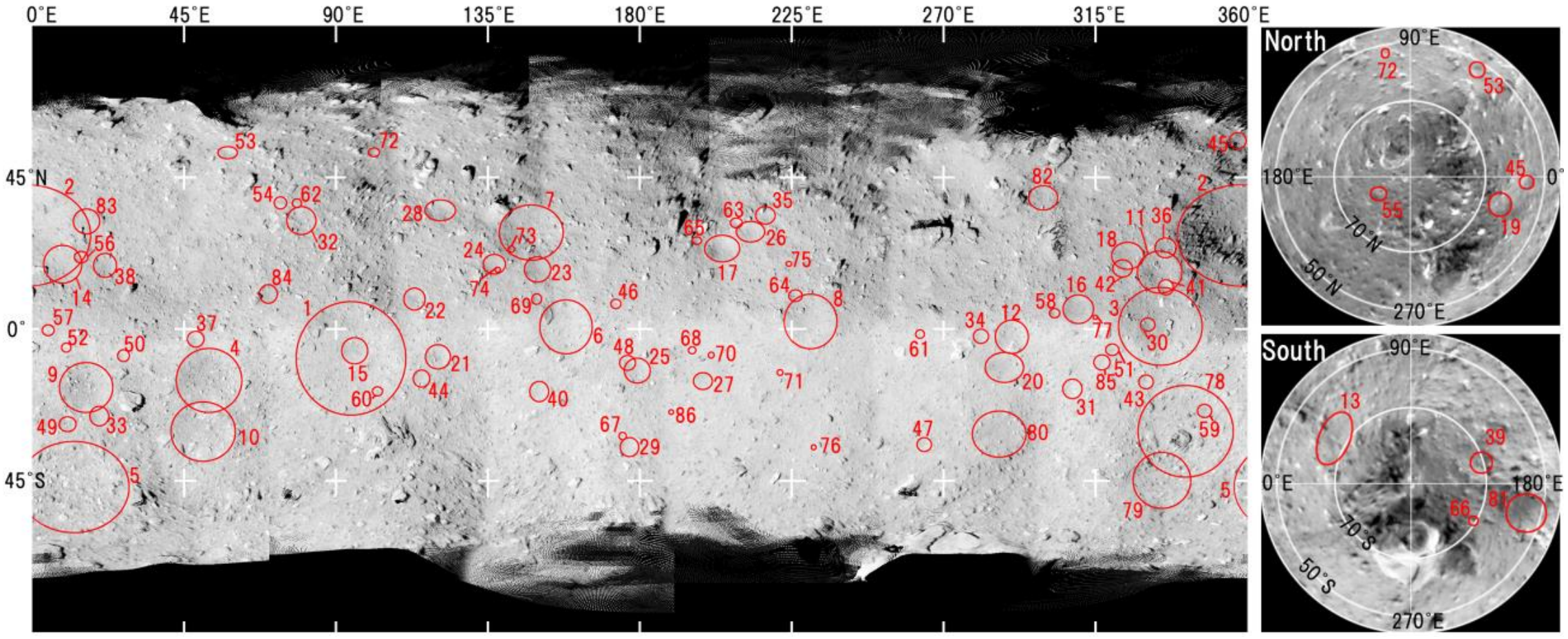

**Figure 2.** The distribution of craters on Ryugu. Red lines roughly outline the rim of craters. On the left is a simple cylindrical projection mosaic map derived from the July 20, 2018 images. On the right is the azimuthal equidistant projection maps centered in the north (top) and south poles (bottom) derived from January 24, 2019 and August 23, 2018. Numbers correspond to Figure 1 and Table 3.

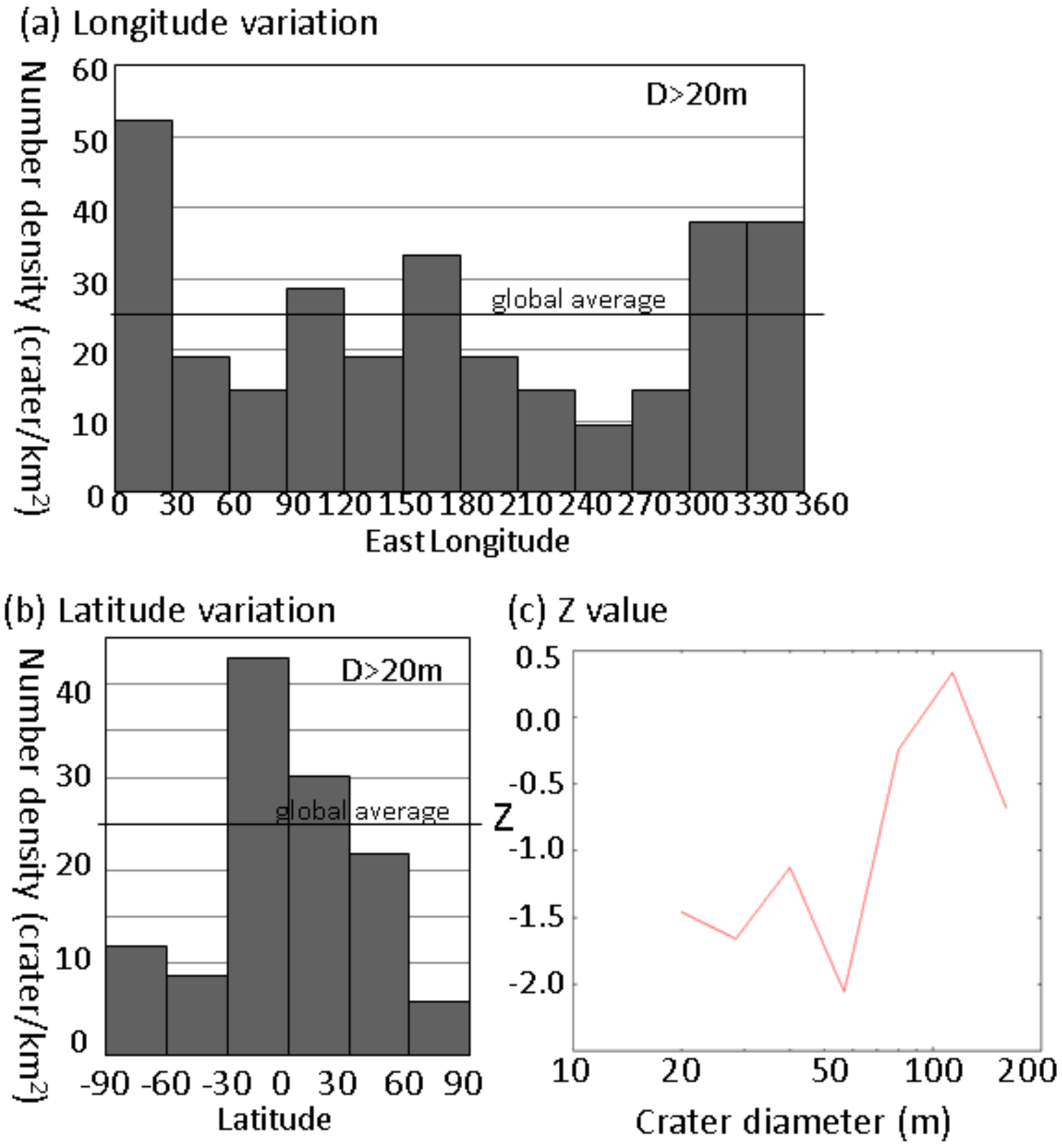

**Figure 3.** **(a)** The longitudinal variation in crater density (D ≥ 20 m). **(b)** The latitudinal variation in crater density (D ≥ 20 m). **(c)** The Z value for spatial distribution on Ryugu.

**Table 1. Images utilized in this work.**

| | Date | Num. | Resolution | Note |
|---|---|---|---|---|
| (i) | Jul. 20 2018 | 96 | 0.72 m/px | Low latitude < 70 degree |
| (ii) | Aug.1 2018 | 85 | 0.69-0.55 m/px | Low latitude < 40 degree |
| (iii) | Oct. 4 2018 | 11 | 0.32 m/px | North pole |
| (iv) | Oct. 30 2018 | 18 | 0.62 m/px | North and South poles |
| (v) | Feb. 28 2019 | 26 | 0.67 m/px | North and South poles |
| (vi) | Aug. 23 2018 | 52 | 2.6 m/px | South pole at small emission |
| (vii) | Jan. 24 2019 | 52 | 2.1 m/px | North pole at small emission |

**Table 2.** Classification in this study.

| Classification | Characteristic | Our Judge |
|---|---|---|
| I | Circular depression with rim | Crater |
| II | Circular depression without rim | Crater |
| III | Quasi-circular depression | Crater |
| IV | Quasi-circular features | Not crater |

**Table 3.** Impact craters on Ryugu

| #*1 | Lat. | Lon. (°E) | D (m) | CL*2 |
|---|---|---|---|---|
| *Classification I-III* | | | | |
| 1 | -7.19 | 92.99 | 290 | I |
| 2 | 28.34 | 353.68 | 224 | II |
| 3 | -0.70 | 330.28 | 221 | II |
| 4 | -14.83 | 51.20 | 183 | I |
| 5 | -50.56 | 9.84 | 173 | III |
| 6 | 0.42 | 157.84 | 154 | II |
| 7 | 30.37 | 145.99 | 145 | II |
| 8 | 3.24 | 229.95 | 142 | I |
| 9 | -17.19 | 14.29 | 133 | II |
| 10 | -31.50 | 47.26 | 131 | I |
| 11 | 17.02 | 332.23 | 100 | I |
| 12 | -1.88 | 289.49 | 90.2 | II |
| 13 | -63.73 | 24.45 | 85.0 | III |
| 14 | 16.90 | 6.44 | 79.0 | III |
| 15 | -4.54 | 95.52 | 78.9 | III |
| 16 | 6.01 | 308.75 | 77.1 | I |
| **17** | 23.70 | 205.33 | 76.2 | II |
| **18** | 20.86 | 322.68 | 73.9 | II |
| **19** | 57.73 | 342.98 | 73.3 | II |
| **20** | -11.64 | 287.71 | 69.0 | III |
| **21** | -8.44 | 119.53 | 69.0 | I |
| **22** | 8.28 | 111.52 | 65.8 | III |
| **23** | 18.59 | 149.73 | 65.8 | II |
| **24** | 19.65 | 136.94 | 62.1 | II |
| **25** | -12.18 | 179.11 | 61.0 | II |
| **26** | 28.48 | 213.16 | 58.2 | II |
| **27** | -16.28 | 199.13 | 53.6 | II |
| **28** | 36.81 | 121.83 | 52.2 | III |
| **29** | -36.18 | 176.65 | 51.3 | III |
| **30** | -0.07 | 329.24 | 51.3 | II |
| **31** | -17.20 | 307.19 | 48.8 | II |
| **32** | 33.09 | 81.45 | 48.4 | III |
| **33** | -26.28 | 17.83 | 46.6 | II |
| **34** | -1.87 | 279.89 | 44.2 | II |
| **35** | 33.50 | 217.25 | 44.1 | III |
| **36** | 24.37 | 335.50 | 43.1 | III |
| **37** | -3.08 | 47.02 | 42.7 | I |
| **38** | 18.62 | 20.92 | 42.5 | III |
| **39** | -69.03 | 170.39 | 41.7 | III |
| **40** | -19.49 | 150.94 | 41.4 | III |
| **41** | 11.91 | 334.53 | 39.9 | I |
| **42** | 17.52 | 323.27 | 39.7 | II |
| **43** | -15.38 | 328.28 | 37.2 | II |
| **44** | -14.87 | 115.35 | 36.5 | II |
| **45** | 55.75 | 357.50 | 36.0 | II |
| **46** | 7.79 | 173.00 | 34.6 | I |
| **47** | -34.51 | 263.82 | 34.4 | I |
| **48** | -10.68 | 176.31 | 32.0 | II |
| **49** | -28.57 | 9.83 | 30.6 | III |
| **50** | -7.92 | 26.75 | 29.3 | III |
| **51** | -6.30 | 319.67 | 28.1 | III |
| **52** | -5.81 | 9.76 | 27.6 | III |
| **53** | 52.36 | 57.31 | 27.5 | II |
| **54** | 37.10 | 73.44 | 26.5 | III |
| **55** | 79.00 | 209.00 | 26.0 | II |
| **56** | 21.44 | 13.71 | 25.3 | III |
| **57** | -0.89 | 4.52 | 24.4 | III |
| **58** | 4.61 | 301.74 | 23.6 | I |
| **59** | -24.00 | 346.55 | 21.7 | III |
| **60** | -19.26 | 102.14 | 21.5 | III |
| **61** | -2.08 | 263.10 | 21.0 | II |
| **62** | 37.02 | 78.22 | 20.9 | III |
| **63** | 31.21 | 208.26 | 20.2 | III |
| **64** | 8.40 | 225.72 | 17.2 | I |
| **65** | 25.81 | 196.97 | 16.2 | II |
| **66** | -69.00 | 211.00 | 16.0 | II |
| **67** | -31.59 | 174.66 | 14.5 | II |
| **68** | -6.96 | 195.40 | 14.4 | III |
| **69** | 9.65 | 149.10 | 14.0 | I |
| **70** | -8.37 | 201.10 | 14.0 | II |
| **71** | -12.89 | 221.27 | 13.9 | II |
| **72** | 52.43 | 100.93 | 12.6 | II |
| **73** | 23.18 | 141.89 | 12.5 | II |
| **74** | 17.15 | 137.76 | 11.4 | II |
| **75** | 19.01 | 224.05 | 11.2 | II |
| **76** | -35.88 | 231.29 | 10.5 | II |
| **77** | 3.40 | 314.61 | 10.0 | II |
| ***Classification IV*** | | | | |
| **78** | -31.21 | 341.74 | 223 | IV |
| **79** | -45.92 | 334.99 | 133 | IV |
| **80** | -30.64 | 284.23 | 112 | IV |
| **81** | -58.37 | 198.43 | 112 | IV |
| **82** | 38.21 | 298.88 | 55.8 | IV |
| **83** | 31.87 | 15.43 | 53.0 | IV |
| **84** | 10.26 | 69.87 | 46.7 | IV |
| **85** | -10.59 | 316.99 | 35.0 | IV |
| **86** | -25.36 | 189.21 | 11.4 | IV |

*1 Number in descending order of diameter. Ryugu has 7 named craters:

Urashima (No.1), Cendrillon (No.2), Kolobok (No.3), Momotaro (No.4), Kintaro (No. 6), Brabo (No. 8), and Kibidango (No.10).

*2 Classification shown in Table 2.

**References**

Güttler, C., N. Hirata, and A.M. Nakamura (2012) Cratering experiments on the self armoring of coarse-grained granular targets. *Icarus* 220(2), 1040-1049.

Hirabayashi et al. (2019), The Western Bulge of 162173 Ryugu Formed as a Result of a Rotationally Driven Deformation Process. *The Astrophysical Journal Letters* 874, Number 1.

Hirata N. (2017) Spatial distribution of impact craters on Deimos. *Icarus* 288, 69-77.

Jacobson S. A. and D. J.Scheeres (2011) Dynamics of rotationally fissioned asteroids: Source of observed small asteroid systems. *Icarus* 214(1), 161-178.

Kahn, E. G., O. S. Barnouin, D. L. Buczkowski, C. M. Ernst, N. Izenberg, S. Murchie and L. M. Prockter. (2011) A Tool for the Visualization of Small Body Data. In *42nd Lunar and Planetary Science Conference*, Abstract #1618. Houston: Lunar and Planetary Institute.

Lissauer, J. J., S. W. Squyres, and W. K. Hartmann (1988), Bombardment History of the Saturn System, Journal of Geophysical Research: Solid Earth, 93(B11), 13776-13804.

Michel, P., D. P. O'Brien, S. Abe, and N. Hirata, (2009) Itokawa's cratering record as observed by Hayabusa: implications for its age and collisional history. *Icarus* 200, 503–513.

Phillips, R. J., R. F. Raubertas, R. E. Arvidson, I. C. Sarkar, R. R. Herrick, N. Izenberg and R. E. Grimm (1992) Impact craters and Venus resurfacing history. *Journal of Geophysical Research: Planets* 97, 15923-15948.

Robinson, M. S., P. C. Thomas, J. Veverka, S. L. Murchie, and B. B. Wilcox (2002) The geology of 433 Eros. *Meteoritics & Planetary Science* 37(12), 1651-1684.

Scheeres, D.J., (2015) Landslides and Mass shedding on spinning spheroidal asteroids. *Icarus* 247, 1-17.

Squyres, S. W., C. Howell, M. C. Liu and J. J. Lissauer (1997) Investigation

of Crater "Saturation" Using Spatial Statistics. *Icarus* 125, 67-82.

Sugita, S. et al. (2019) The geomorphology, color, and thermal properties of Ryugu: Implications for parent-body processes. *Science* 364, 272-275.

Thomas, P. C. and M.S. Robinson, (2005) Seismic resurfacing by a single impact on the asteroid 433 Eros. *Nature* 436, 366–369.

Walsh, K. J. et al. (2019), Craters, boulders and regolith of (101955) Bennu indicative of an old and dynamic surface. *Nature Geoscience* 12, 242–246.

Watanabe S. et al. (2019) Hayabusa2 arrives at the carbonaceous asteroid 162173 Ryugu—A spinning top–shaped rubble pile. *Science* 364, 268-272.